%% file: Moravec_MOO_1506_arXiv.tex
\documentclass[twocolumn,twocolappendix]{aastex63}
\usepackage{amsmath}	% Advanced maths commands
\usepackage{amssymb}	% Extra maths symbols
\usepackage{wasysym} % Extra maths symbols
\usepackage{ifsym}
\usepackage{pifont}
\usepackage{stix}
\usepackage{multirow,tabularx}
\usepackage{subfigure}
\usepackage{booktabs}
\usepackage{relsize}
\usepackage{color}  
\usepackage{xcolor}
\usepackage[hyphenbreaks]{breakurl}
\usepackage[hyphens]{url}
\def\wise{\textit{WISE}}
\def\madcows{MaDCoWS}
\def\spz{\textit{Spitzer}}
\def\as{$^{\prime\prime}$}
\def\am{$^{\prime}$}
\def\m2{MUSTANG-2}
\def\dc{DECaLS}
\def\d{$^{\circ}$}

\def\moo15{MOO J1506+5137}
\def\deg{$^{\circ}$}
\received{March 20, 2020}
\revised{May 14, 2020}
\accepted{\today}
\submitjournal{ApJ}

\shorttitle{High Radio Activity in a Merging Cluster}
\shortauthors{E. Moravec et al.}

\graphicspath{{./}{figures/pdf/}}
\begin{document}

\title{The Massive and Distant Clusters of \wise\ Survey IX: High Radio Activity in a Merging Cluster}

\correspondingauthor{Emily Moravec}
\email{emily.moravec@asu.cas.edu}
\input{authors.tex}

%%%%%%%%%%%%%%%%%%%%%%%%%%%%%%%%%%%%%
% Abstract of the paper
\begin{abstract}
We present a multi-wavelength investigation of the radio galaxy population in the galaxy cluster MOO J1506+5137 at $z$=1.09$\pm$0.03, which in previous work we identified as having multiple complex radio sources. The combined dataset used in this work includes data from the Low-Frequency Array Two-metre Sky Survey (LoTSS), NSF's Karl G. Jansky Very Large Array (VLA), the Robert C. Byrd Green Bank Telescope (GBT), the \textit{Spitzer Space Telescope}, and the Dark Energy Camera Legacy Survey (\dc). We find that there are five radio sources which are all located within 500 kpc ($\sim$1\am) of the cluster center and have radio luminosities $P_{\mathrm{1.4GHz}}$ > 1.6$\times$10$^{24}$ W Hz$^{-1}$. The typical host galaxies are among the highest stellar mass galaxies in the cluster. The exceptional radio activity among the massive galaxy population appears to be linked to the dynamical state of the cluster. The galaxy distribution suggests an ongoing merger, with a subgroup found to the northwest of the main cluster. Further, two of the five sources are classified as bent-tail sources with one being a potential wide-angle tail (WAT)/hybrid morphology radio source (HyMoRS) indicating a dynamic environment. The cluster also lies in a region of the mass-richness plane occupied by other merging clusters in the Massive and Distant Clusters of \wise\ Survey (\madcows). The data suggest that during the merger phase radio activity can be dramatically enhanced, which would contribute to the observed trend of increased radio activity in clusters with increasing redshift.
\end{abstract}


%%%%%%%%%%%%%%%%%%%%%%%%%%%%%%%%%%%%%%%%%%%%%%%%%%
%%%%%%%%%%%%%%%%% BODY OF PAPER %%%%%%%%%%%%%%%%%%
\section{Introduction}\label{intro}
Active galactic nuclei (AGNs) that emit radio synchrotron emission (i.e., radio galaxies, radio-AGN) display a classical double-lobed morphology and are typically categorized as Fararoff-Riley type I (FR I) or Fanaroff-Riley type II (FR II) \citep[see][]{FR74}. FR I sources are `edge-darkened' meaning that the emission is brighter near the radio core and becomes fainter radially outward. FR II sources are `edge-brightened' meaning that the well-separated lobes contain are distinctive areas of brightest emission (i.e., `hotspots') near the edge of the lobe. Often radio galaxies in clusters will show signs of interaction with the dense intracluster medium (ICM) found in clusters \citep{Miley72,Blanton00,PM17,Garon19}. In this situation, the radio lobes will become bent and distorted due to ram pressure producing `bent-tail' sources \citep{Miley72,Burns98}. 

There is evidence for heightened radio-AGN activity in the cluster environment which is showcased in a variety of ways \citep{Miller03,Wylezalek13,PM17,Mo18}. The association of radio galaxies with galaxy clusters dates back to the 1950s \citep{Baade54}. The relationship between radio galaxies and galaxy clusters is so strong that radio-AGN have been used to successfully identify rich, high redshift ($z\gtrsim1$) clusters, as well as $z\gtrsim2$ protoclusters: the Clusters Around Radio-Loud AGN program \citep[CARLA,][]{Wylezalek13,Wylezalek14,Noirot16,Noirot18}, the Clusters Occupied by Bent Radio AGN \citep[COBRA,][]{PM17}, \cite{Castignani14}, and \cite{Rigby14}. 
More recently, \cite{Mo18} find heightened activity of radio-AGN within 1\am\ of the cluster centers and Mo et al. (sub.) find that this activity that is within 0.5 Mpc increases strongly as a function of redshift. And more specifically it has been shown that merging clusters can foster radio activity in powerful AGNs \citep{Miller03} and optically faint star-forming galaxies \citep{Owen99,Miller03}.

Dense environments are known to affect the radio morphology of the radio galaxies within them. For the canonical double-lobed radio galaxies, FR I sources are found in richer environments than FR II sources \citep{Longair79,prestage88,Owen99,Miller99,Wing11,Gendre13,Croston19}. Additionally, rare radio galaxies called Hybrid Morphology Radio Sources \citep[$<$1\%;][]{Gawronski06} have been discovered that display FR I radio structures on one side of the nucleus and FR II on the other. These anomalous radio galaxies could be explained as a product of environmental differences \citep[HyMoRS:][]{Gopal-Krishna00,Kapinska17}. 

In a uniform environment with a line of sight perpendicular to the jets, double-lobed radio galaxies canonically display symmetric jets and lobes. However, as radio galaxies travel through the dense ICM of galaxy clusters, the ram pressure due to the relative motion of the radio galaxy host can cause the lobes to bend opposite the direction of motion of the galaxy producing a bent-tail morphology (BTs hereafter). There are several types of BTs that are classified according to their morphologies: narrow-angle tail (NAT), head-tail (HT), and wide-angle tail (WAT) radio galaxies. NATs result from the radio jets being swept back by ram pressure ($P_{\rm ram} = \rho v^2$) as the host galaxy moves at high speed through the ICM \citep{Miley72,Begelman79}. The higher the velocity of the galaxy, the higher the bending angle \citep{Miley72,Begelman79}. NATs typically subtend small angles and the most extreme case is where the tails become indistinguishable from one another, forming a head-tail radio galaxy \citep{Miller03,Gregory17}. 

WATs are fundamentally different from NATs in a variety of ways. A WAT is a source with twin, well-collimated jets that suddenly flare into plumes, which are different from normal smoothly expanding FR I jets \citep{Leahy93}. The direction of these plumes can either deviate from the jet axis and become bent or align with the jet axis \citep{Hardcastle98,Hardcastle04}. When bent, the opening angle is typically large and the result is a `C-shaped' source. One of the main differences from NATs is that WAT host galaxies are cluster-center objects and are thus near the bottom of the cluster potential. Therefore, WAT host galaxies must be nearly at rest with respect to the cluster gravitational potential \citep{Owen76,Quintana82,Eilek84}. Thus, any bent WAT morphology cannot be explained by ram pressure from the motion of the host galaxy as it is for NATs. Instead it has been proposed that the bending of a WAT is an indication of large-scale bulk motions in the ICM and can thus be indicative of a cluster merger \citep{Roettiger93,Pinkney94,Burns98,Sakelliou00}.

Structure in the Universe is viewed as evolving hierarchically, with large features such as clusters forming through the repeated mergers of smaller groups \citep[e.g.,][]{Evrard90,Jing95,Frenk96,BK09,Pillepich18}. The cluster merging process is chaotic in nature as kinetic energy of the colliding subclusters dissipates in the intracluster gas through shock heating, giving rise to strong, spatial variations of gas temperature and entropy, as well as gas bulk flows, destruction of cooling flows, turbulence, and thermal conduction within the ICM (e.g., the Bullet Cluster: \citealt{Markevitch02}, `El Gordo': \citealt{Menanteau12}, Abell 2255: \citealt{Miller03}, Abell 2142: \citealt{Markevitch00}). Clusters are still forming and rapidly assembling \citep{Bode01} during the $1<z<2$ epoch, which is thus an important and transformative period in the early Universe to investigate the connection between environment and cluster galaxy evolution.

In this work, we investigate a high-redshift cluster, MOO J1506+5137, that has a high number of radio sources near its center and shows evidence for being a merging system. In Section \ref{M1506}, we describe the properties of the cluster. In Section \ref{sect:obs}, we provide an overview of the rich, multi-wavelength dataset collected for this cluster and the details of our reduction. In Section \ref{sect:radio_morph}, we describe and categorize the radio sources and in Section \ref{sect:cps} we discuss their infrared and optical counterparts. In Section \ref{sect:bcgs}, we specifically discuss the candidates for brightest cluster galaxy (BCG) in the system. In Section \ref{sect:merger}, we discuss evidence that this cluster is a merger. Lastly, in Section \ref{sect:discuss} we discuss possible causes of the high radio activity of this cluster. Throughout this work, we adopt the flat $\Lambda$CDM cosmological model with a \cite{Planck15} cosmology, $H_0$ = 67.8 km s$^{-1}$, $\Omega_{m}$ = 0.308, $\Omega_{\Lambda}$ = 0.692, and $n_s$ = 0.968. Unless otherwise noted, all \spz\ magnitudes are on the Vega system and \dc\ magnitudes are on the AB system in order to remain on the fiducial system of the surveys\footnote{The conversion from Vega to AB for [3.6] is [3.6]$_{\mathrm{Vega}}$ = [3.6]$_{\mathrm{AB}}$ - 2.79 and for [4.5] the conversion is [4.5]$_{\mathrm{Vega}}$ = [4.5]$_{\mathrm{AB}}$ - 3.26}.

\begin{deluxetable}{cccccc}
	\tablecaption{MOO J1506+5137 Properties \label{tb:info}}
	\tablehead{\colhead{R.A.} & \colhead{Dec.} & \colhead{$z_{\mathrm{phot}}$}  & \colhead{$\lambda_{15}$} & \colhead{M$_{500}$}\\
	 & & & & (10$^{14}$M$_{\odot}$)}
    \startdata
    \smallskip
	 15:06:20.7 & 51:37:01 & 1.09$\pm$0.03 & 74$\pm$8 & 3.17$\pm0.29$\\
    \enddata
	\tablecomments{The R.A. and Dec. are from the \madcows\ cluster search using \wise\ data. $z_{\mathrm{phot}}$ is the photometric redshift determined using a combination of Pan-STARRS data and \spz\ data (see \S\ref{sect:spz}). $\lambda_{15}$ is the richness calculated using \spz\ 4.5$\mu$m data. M$_{500}$ is calculated through profile fits to the GBT data and the error is statistical (see Dicker et al. in prep for more details).}
\end{deluxetable}

\section{The Curious Case of MOO J1506+5137}\label{M1506}
MOO J1506+5137 is a high-redshift ($z_{\mathrm{phot}}$ = 1.09$\pm$0.03) massive ($M_{500}$ = 3.17$\pm{0.29}$ $\times$ 10$^{14}$ M$_{\odot}$) cluster (See Table \ref{tb:info}) identified in The Massive and Distant Clusters of WISE Survey (\madcows: \citealt{Gonzalez19}). \madcows\ identifies cluster candidates at 0.8 $\lesssim z \lesssim$ 1.4 by cross-matching the \textit{Wide-field Infrared Survey Explorer} \citep[\textit{WISE}:][]{Wright10} with the Panoramic Survey Telescope and Rapid Response System \citep[Pan-STARRS:][]{PS16} and applying magnitude and color cuts to isolate cluster galaxies. From a smoothed density map of the cluster galaxies, \cite{Gonzalez19} identified the 2681 highest amplitude detections. The primary \madcows\ search covers 17,668 deg$^2$ of the extragalactic sky at $\delta$ $>$ $-$30$^{\circ}$ providing the largest survey of $z>$1 clusters independent from Sunyaev-Zel'dovich (SZ) effect measurements making \madcows\ the widest field survey at this epoch. The cluster center reported in Table \ref{tb:info} is the catalog coordinates from this original \textit{WISE}---PanSTARRS search and has a positional uncertainty of 21\as (see \citealt{Gonzalez19} for a more detailed discussion).

MOO J1506+5137 stands out in the \madcows\ sample due to its high radio activity. From the 1300 highest significance \madcows\ clusters in the Karl G. Jansky Very Large Array (VLA) Faint Images of the Radio Sky at Twenty-cm survey (FIRST: \citealt{Becker94}) footprint, we identified a sample of 51 clusters with extended radio sources defined as having at least one FIRST source with a deconvolved size exceeding 6\farcs5 ($\sim$50 kpc at z$\sim$1) within 1\am\ ($\sim$500 kpc at z$\sim$1) of the cluster center. This sample was observed with the VLA as a part of a larger study \citep{Moravec20}. 

Through these VLA follow-up observations, we discovered that MOO J1506+5137 is unique in several ways in this sample of \madcows\ clusters that contain extended radio sources. Pertaining to the number of FIRST sources, \moo15\ is one of the eight clusters that have three FIRST sources within 500 kpc ($\sim$1\am) and is one of two clusters that have four or more FIRST sources within 500 kpc of the cluster center. Similarly in the higher resolution and deeper VLA data, the most radio sources above a 4$\sigma$ detection that any cluster has within 500 kpc of the cluster center is five, and \moo15\ is one of two such clusters. Next, the highest number of complex sources (defined as having non-point-like source structure) within 500 kpc of the cluster center that any cluster has is three, and \moo15\ is again one of only two such clusters. Most importantly, \moo15\ is the only cluster that has three complex sources that have jet-like structures. Lastly, it is the only cluster that contains two bent-tail sources within 500 kpc of the cluster center. In summary, \moo15\ is set apart from other \madcows\ clusters and radio-active \madcows\ clusters with its high number of radio sources (5) and the number of complex radio sources that appear to be AGN jets (3), of which two are bent tail sources (see Figure \ref{fig:radio_data}). We note that these statistics concerning the sources in the VLA data exclude clusters containing low redshift interloper radio galaxies (see \citealt{Moravec20}).

\begin{deluxetable*}{ccccc}
    \tablecaption{Radio Observations\label{tb:obs}} 
    \tablehead{\colhead{$\nu$} & \colhead{Telescope/Survey} & \colhead{Date} & \colhead{Res.} & \colhead{RMS}}
    \startdata
	 144 MHz & LoTSS & DR1 & $\sim$6\as & 0.135 mJy/beam\\
     1.4 GHz & JVLA & 08 May 2018 & 1\farcs2 $\times$ 1\farcs1 & 23 $\mu$Jy/beam\\
     6.0 GHz & JVLA & 13 Oct. 2017 & 1\farcs0 $\times$ 0\farcs8 & 8 $\mu$Jy/beam\\
     90 GHz & GBT & 2019 \& 2020 & $\sim$10\as & 36 uK/beam\\
     \enddata
\end{deluxetable*}

\section{Observations} \label{sect:obs}
To investigate MOO J1506+5137, we assembled the suite of multi-wavelength observations (see Table \ref{tb:obs}) detailed below: 
\begin{itemize}
    \item Low-Frequency Array (LOFAR) 144 MHz
    \item VLA 1.4 GHz
    \item VLA 6.0 GHz
    \item Robert C. Byrd Green Bank Telescope (GBT)\footnote{The Green Bank Observatory is a major facility supported by the National Science Foundation and operated under cooperative agreement by Associated Universities, Inc.} 90 GHz
    \item \spz\ IRAC 3.6 and 4.5 $\mu$m
    \item Dark Energy Camera (DECam) $r$ \& $z$
\end{itemize}
Each of these observations provides important information about either the radio sources, their counterparts, or the galaxy cluster properties. The intermediate-frequency radio observations (1.4 and 6.0 GHz) trace the current energy injection through the jets and hotspots, while low-frequency radio observations (144 MHz) allow exploration of the full extent of the jet emission and the history. Both facilitate radio source morphological classification and analysis (see \S\ref{sect:radio_morph}). The GBT observations provide a mass estimate for the cluster through the SZ effect and also high-frequency flux densities for some of the radio sources (see \S\ref{sect:GBT}). Finally, the \spz\ and \dc\ data allow stellar mass and membership determination.

\subsection{LoTSS Data} \label{sect:LTS}
The LOFAR Two-metre Sky Survey (LoTSS) is an ongoing sensitive, high-resolution 120-168 MHz survey of the entire northern sky\footnote{https://www.lofar-surveys.org/releases.html}. In February 2019, the first full-quality public data release of LoTSS became available (DR1), presenting  2\% of the eventual coverage \citep{Shimwell19}. Observations of MOO J1506+5137 were included in this first data release. The median sensitivity is S$_{\mathrm{144 MHz}}$ = 71 $\mu$Jy beam$^{-1}$ and the resolution of the images is 6$^{\prime\prime}$. For MOO J1506+5137, we determine an RMS near the source of 1.35$\times$ $10^{-4}$ Jy/beam.

\subsection{VLA Observations, Image Processing, and Flux Densities}\label{sect:vla}
VLA high-resolution data were taken during 17B (PI: Gonzalez, 17B-197), and 18A (PI: Moravec, 18A-039). The data taken in 17B were taken in C-Band in the B configuration. The observations were centered at 5.5 GHz with a bandwidth of 1.9 GHz. Given this configuration and band, the angular resolution was 1\farcs0 $\times$ 0\farcs8\ and the primary beam full-width half power (FWHP) was 8$^{\prime}$. During 17B, the cluster was observed three times for nine minutes each within a scheduling block for a total of $\sim$28 minutes on source. 

The data taken during 18A were taken in L-Band in the A configuration. The observations were centered at 1.4 GHz (21 cm) with a bandwidth of 600 MHz. Given this configuration and band, the angular resolution was 1\farcs2 $\times$ 1\farcs1 and the primary beam full-width half power (FWHP) was 30$^{\prime}$. The cluster was observed twice for six minutes within a scheduling block for a total of $\sim$12 minutes on source. For all observations the correlator was configured with 16 spectral windows, each with 64 channels.

The data were flagged, calibrated, and imaged with Common Astronomy Software Applications (CASA) package versions $>$ 5.0 \citep[]{casa}. All measurement sets were first processed through the VLA CASA Calibration Pipeline for basic flagging and calibration. We created images by applying the \texttt{tclean}\footnote{https://casa.nrao.edu/casadocs-devel/stable/global-task-list/task\_tclean/about} algorithm. We used a pixel scale of 0.28$^{\prime\prime}$ for L-Band and 0.24$^{\prime\prime}$ for C-Band, \texttt{specmode=}`mfs', and \texttt{weighting=}BRIGGS (robust=0.5). In both cases, we performed several rounds of phase-only self-calibration to increase the S/N ratio, reduce the prominence of improper cleaning artifacts, and recover more of the source structure.

The RMS is $\sim$23 $\mu$Jy per beam for the L-Band image and $\sim$8 $\mu$Jy per beam for the C-Band image. The RMS values were determined by the following process. Using the CASA viewer, we calculated the RMS individually in $\sim$ 4 square regions that were free of any sources and whose locations were chosen to sample the full area near the targeted source. These individual RMS measurements were averaged together to produce the final RMS.

The R.A. and Dec. of the radio sources reported in Table \ref{tb:rad_sources} are the inferred origin of the radio emission based on the morphological classification, determined using the C-Band data. The 1.4 and 6.0 GHz total flux densities were calculated for each of the radio sources using the CASA viewer. Regions were drawn to follow the 4$\sigma$ contours and the flux densities within these contours are reported in Table \ref{tb:rad_sources}. The errors are $\sqrt{N_{\mathrm{beams}}}\sigma_i$ where $N_{\mathrm{beams}}$ is the number of beams contained within the region in which the flux density is calculated and $\sigma_i$ is the RMS of the image in Jy/beam. We calculate the radio power (radio luminosity) in Table \ref{tb:rad_sources},
\begin{equation}\label{eqn:power}
    P_{1.4} = 4\pi{D_L}^2S_{\mathrm{1.4GHz}}(1+z)^{\alpha-1},
\end{equation}
where $D_L$ is the luminosity distance at the photometric redshift of the cluster ($z$=1.09, see Table \ref{tb:info}), $z$ is the photometric redshift of the cluster, $S_{1.4}$ is the integrated radio flux density at 1.4 GHz from the VLA image, the (1 + $z$)$^{\alpha-1}$ term includes both the distance dimming and K-correction where $\alpha$ is the radio spectral index ($S_{\nu} \propto \nu^{-\alpha}$). Typical values of $\alpha$ for extended radio sources range from 0.7 to 0.8 \citep[]{Kellermann88,Condon92,PetersonBook,L&M07,Miley08,Tiwari19} and we adopt $\alpha$=0.8 as in \citet[]{Chiaberge09}, \citet[]{Gralla11}, and \citet[]{Yuan16}. The lowest power source within $\sim$75\as\ of the cluster center that we detect is $P_{1.4}$ = 1.6$\times$10$^{24}$ W Hz$^{-1}$ (see Table \ref{tb:rad_sources}).

\begin{figure*}
    \centering
	\includegraphics[width=\linewidth, keepaspectratio]{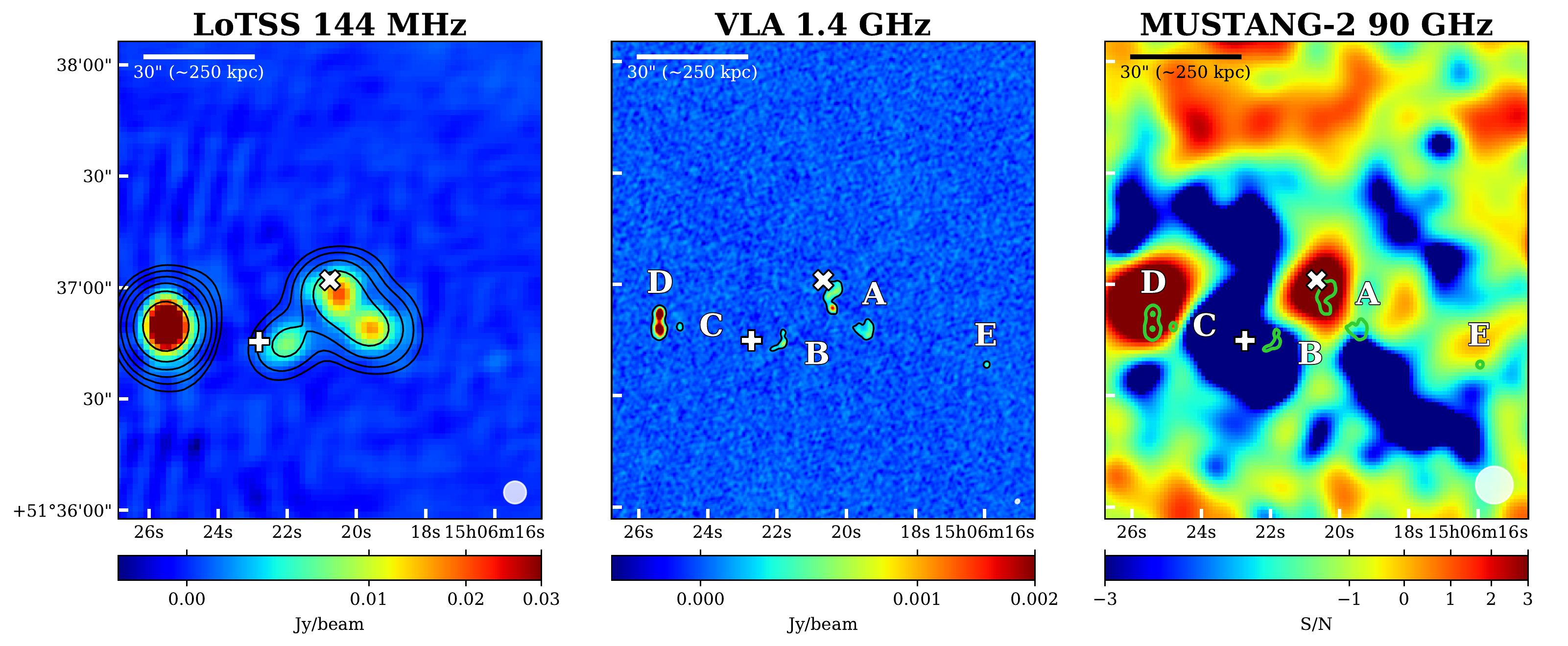}
	\includegraphics[width=\linewidth, keepaspectratio]{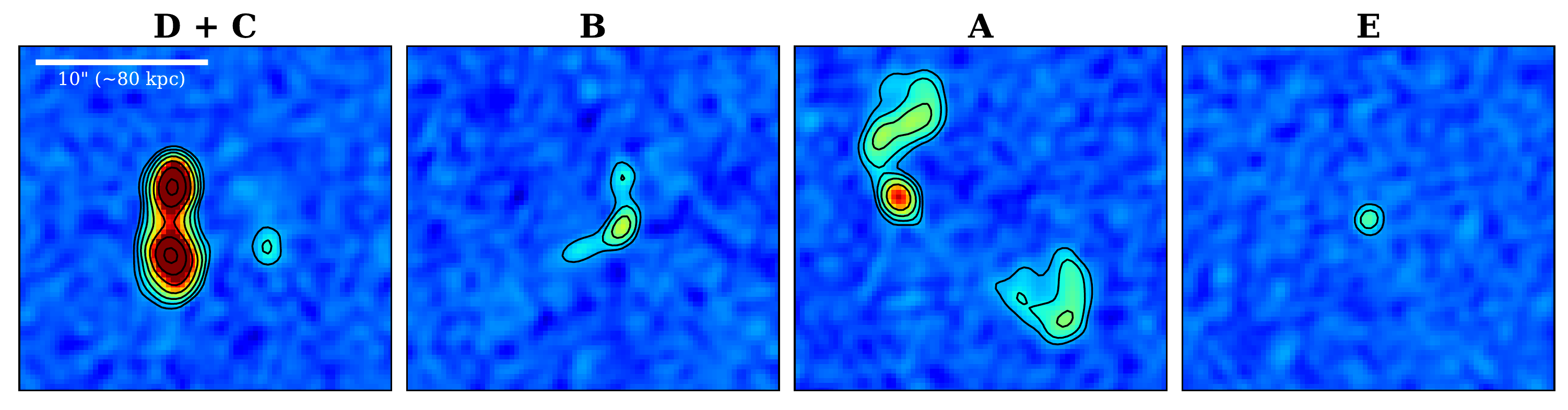}
	\caption{\textbf{Top row:} 2.0\am\ $\times$ 2.0\am\ ($\sim$975 $\times$ $\sim$975 kpc) radio images of MOO J1506+5137. The source labels correspond to the radio sources listed in Table \ref{tb:rad_sources}. \textit{Upper left:} 144 MHz LoTSS data where the contour levels start at 4$\sigma$ and increase by factors of 2$^n$ where n = 1,2,3, etc. The contours are smoothed by a Gaussian 3x3 pixel smoothing kernel. \textit{Upper middle:} 1.4 GHz (L-band) VLA data with a contour level of 4$\sigma$ shown in black. \textit{Upper right:} 1.4 GHz VLA 4$\sigma$ and 256$\sigma$ (green) contours overlaid on the 90 GHz \m2\ signal-to-noise ratio image. The cluster SZ decrement is clearly seen. For all, the synthesized beam size is shown in the lower right hand corner. The SZ center is denoted by a white + and the \madcows\ \textit{WISE} center is denoted by a white x. \textbf{Bottom row:} VLA 1.4 GHz 20\as\ $\times$ 20\as\ cut-outs of the radio sources. The source letter is listed above each cut-out (see Table \ref{tb:rad_sources}). The contour levels start at 4$\sigma$ and increase by factors of 2$^n$ where n = 1,2,3, etc. The intensity scaling for these images in the bottom panels is the same as in the upper middle panel (VLA 1.4 GHz). For all images, north is up and east is to the left. The color scale of each image in this figure uses a square root stretch function.}
	\label{fig:radio_data}
\end{figure*}

\begin{deluxetable*}{cccccccc}
	\tablecaption{Radio Source Properties \label{tb:rad_sources}}
	\tablehead{\colhead{ID} & \colhead{R.A.} & \colhead{Dec.} & \colhead{R$_{\mathrm{cc}}$} & \colhead{Morph.} & \colhead{$S_{1.4}$} & \colhead{$S_{6.0}$} & \colhead{$P_{\mathrm{1.4}}$} \\
	\colhead{} & \colhead{(J2000)} & \colhead{(J2000)} & \colhead{(\as)} & \colhead{} & \colhead{(mJy)} & \colhead{(mJy)} & \colhead{(10$^{25}$ W Hz$^{-1}$)} }
	\startdata
     A &  15:06:20.40 &  +51:36:53.7 &         8.0 &  BT/WAT/HyMoRS &  8.22$\pm$0.13 &   2.86$\pm$0.04 &   4.84$\pm$0.07 \\
     B &  15:06:21.83 &  +51:36:44.2 &        20.0 &          BT &  1.37$\pm$0.06 &   0.44$\pm$0.02 &    0.8$\pm$0.03 \\
     C &  15:06:24.81 &  +51:36:48.8 &        41.0 &          PS &  0.27$\pm$0.03 &   0.06$\pm$0.01 &   0.16$\pm$0.02 \\
     D &  15:06:25.42 &  +51:36:50.1 &        46.0 &       FR II &  33.46$\pm$0.1 &  11.25$\pm$0.04 &  19.71$\pm$0.06 \\
     E &  15:06:15.94 &  +51:36:38.5 &        49.0 &          PS &  0.29$\pm$0.03 &   0.05$\pm$0.01 &   0.17$\pm$0.02 \\
	\enddata
	\tablecomments{ID refers to the letter ID of the corresponding radio source (see Figure \ref{fig:radio_data}). The R.A. and Dec. are the coordinates of the inferred origin of the radio emission based on the morphological classification. R$_{\mathrm{cc}}$ is the distance of the source from the cluster center (defined as the coordinates in Table \ref{tb:info}) in arcseconds. Morph. is the radio morphology of the source based on the VLA data where BT stands for bent-tail, WAT stands for wide-angle tail, HyMoRS stands for hybrid morphology radio source, and PS stands for point source. $S_{1.4}$ is the VLA 1.4 GHz integrated flux density and $S_{6.0}$ is the VLA 6.0 GHz integrated flux density. $P_{1.4}$ is the power of the source at 1.4 GHz calculated according to Eqn. \ref{eqn:power} using the redshift of the cluster ($z=$1.09). The error associated with this calculation is solely statistical and is based on the error in the flux density.\\}
\end{deluxetable*}

\subsection{GBT Observations} \label{sect:GBT}
Over the winters spanning 2018/2019 and 2019/2020, MOO J1506+5137 was observed at 90 GHz with \m2\ on the 100-meter Robert C. Byrd Green Bank Telescope (PI: Brodwin 18B-215 and 19B-200). \m2\ is a 215-element array of feedhorn-coupled Transition Edge Sensor (TES) bolometers which achieves a resolution of $\sim$10\as\ and has an instantaneous field of view of 4.\am25 \citep{Dicker14,Stanchfield16}.

The cluster was observed using Lissajous daisy scans with radii of 2.5\am\ (see \citealt{Romero17}, \citealt{Romero20}, Dicker et al. in prep, and Sievers et al. in prep for an overview of observational techniques and data reduction). A calibration point source was observed every 20-30 minutes to track the telescope pointing and gain. The ALMA grid calibrator J1058+0133 was observed early in the observing run for absolute flux calibration\footnote{https://almascience.eso.org/sc/} \citep{Fomalont14,vanKempen14}. MOO J1506+5137 was observed under project ID AGBT18B\_215, session \#12, for 3.10 hours and under project ID AGBT19B\_200, session \#5 for 2.6 hours, for a total on-source integration time of 5.7 hours. 

The GBT 90 GHz image (see Figure \ref{fig:radio_data}) was produced using the MUSTANG IDL Data Analysis System (MIDAS) which builds off the custom IDL pipeline used with the predecessor of \m2, MUSTANG \citep{Romero15}. Given the telescope scan rate, all sky signal is modulated and lies between 0.1 and 30 Hz, thus frequencies significantly outside this range are filtered out of the raw data. After this a common mode is subtracted to remove atmospheric and readout noise. The data are then gridded to make a 2D map. To estimate the noise the data are divided into two halves and a difference map is made. The SNR map was then made by first smoothing to the MUSTANG-2 beam size, then dividing the signal map by a map of the noise. Because of the common mode removal these MIDAS maps are not unbiased and structures significantly larger than the array will be filtered out of the map. A point source subtracted map was created by identifying the number of point sources and their approximate locations from significant ($>$ 4$\sigma$) peaks in the initial cluster maps. Then, a Gaussian was fit to both the cluster and point sources. With these centers fixed, the point source amplitude and the brightness profile of the cluster were fit. 

The mass is calculated using analysis of the Sunyaev-Zel'dovich effect via the methods described in \citealt{Romero20} and Dicker et al. (in prep.). Broadly, this method has three steps: 
\begin{enumerate}
    \item use a maximum likelihood approach that works directly with the timestreams (MINKASI, Sievers et al. in prep) to fit the cluster center, and the locations and amplitudes of any point sources (of which the number and approximate locations of the point sources are set by hand);
    \item keeping the center fixed, fit for the surface brightness at different radii and point source amplitudes in MINKASI; and
    \item use the surface brightness profile to calculate the mass, assuming a SZ-to-mass conversion \citep{Arnaud10}.
\end{enumerate}
The SZ-based M$_{500}$ estimate using these data is 3.17$\pm{0.29}$ where the quoted error is the statistical error. The SZ center calculated from these maps is $\alpha_{2000}$ = 15h06m22s74, $\delta_{2000}$ = +51\deg36\arcmin44 9\as\ with an error of better than 10\farcs

\subsection{DECaLS Observations} \label{sect:dc}
The Dark Energy Camera Legacy Survey \citep[DECaLS:][]{Dey19} is a product of a survey completed by the Dark Energy Camera (DECam) on the Blanco 4~m telescope, located at the Cerro Tololo Inter-American Observatory. \dc\ is one of the three public projects that, combined, make up the the Dark Energy Spectroscopic Instrument (DESI) Legacy Imaging Surveys\footnote{http://legacysurvey.org/}. For \dc, two-thirds of the DESI footprint was targeted for optical imaging, covering both the North Galactic Cap region at declinations $\leq$ 32$^{\circ}$ and the South Galactic Cap region at declinations $\leq$ 34$^{\circ}$. These data reach 5$\sigma$ depths of $g$=24.0, $r$=23.4 and $z$=22.5 (AB) for a galaxy with a half-light radius of 0\farcs45. Observations began in early 2014 and were completed in March 2019. In this work, we use the $r$ and $z$ magnitudes from DR8 \citep{Dey19}. 

\subsection{Spitzer Observations} \label{sect:spz}
MOO J1506+5137 was observed with \spz\ (PI: Gonzalez, 11080) as part of the larger program to observe $\sim$2000 \madcows\ clusters (90177 and 11080, PI: Gonzalez). For details on the observations and catalog creation see \cite{Moravec20}. The photometric redshift listed in Table \ref{tb:info} is derived from the [3.6]$-$[4.5] and Pan-STARRS $i-$[3.6] colors of galaxies within 1$^{\prime}$ of the cluster location, which are compared with a Flexible Stellar Population Synthesis (FSPS) model \citep{c09,c10}. The richness $\lambda_{15}$ = N$-$N$_{\mathrm{field}}$, where N is the total number of color-selected galaxies within the metric aperture that have fluxes f$_{4.5}>$15$\mu$Jy. Details can be found in \cite{Gonzalez19}.

\section{Radio Source Morphology}\label{sect:radio_morph}
We use both bands of VLA data (1.4 and 6.0 GHz) to classify the morphology of the radio sources (see Figure \ref{fig:radio_data} for images and Table \ref{tb:rad_sources} for the labels). Radio source A is the most central source and is classified first and foremost as a bent-tail (BT) source and secondarily as a WAT/HyMoRS. Both the northern and southern jets contribute to the bent-tail classification. However, the northern jet of the source is significantly bent whereas the morphology of the southern jet is less clear. In the northern jet, there is clear evidence of a bent jet that has an extension northward (possibly a plume) giving it the classification of bent-tail. Combining the C-band data (6.0 GHz), which highlights the direction of the jet, and the L-band data (1.4 GHz), which highlights the extent of the jet, there is evidence that the southern jet is disturbed as well. The emission in the southern lobe is lopsided (toward the east) compared to an undisturbed FR II jet, which would have a undisturbed lobe and relatively equal amounts of material forming each lobe. Because this source has a jet that suddenly flares into the possible beginnings of a plume and it is centrally located in the cluster, this source is classified as a possible WAT. However, higher-resolution radio data are necessary to see whether it has all the features of a classical WAT (e.g. faint well-collimated inner jets and any terminal hotspots).

\begin{figure*}
    \centering
	\includegraphics[width=\linewidth, keepaspectratio]{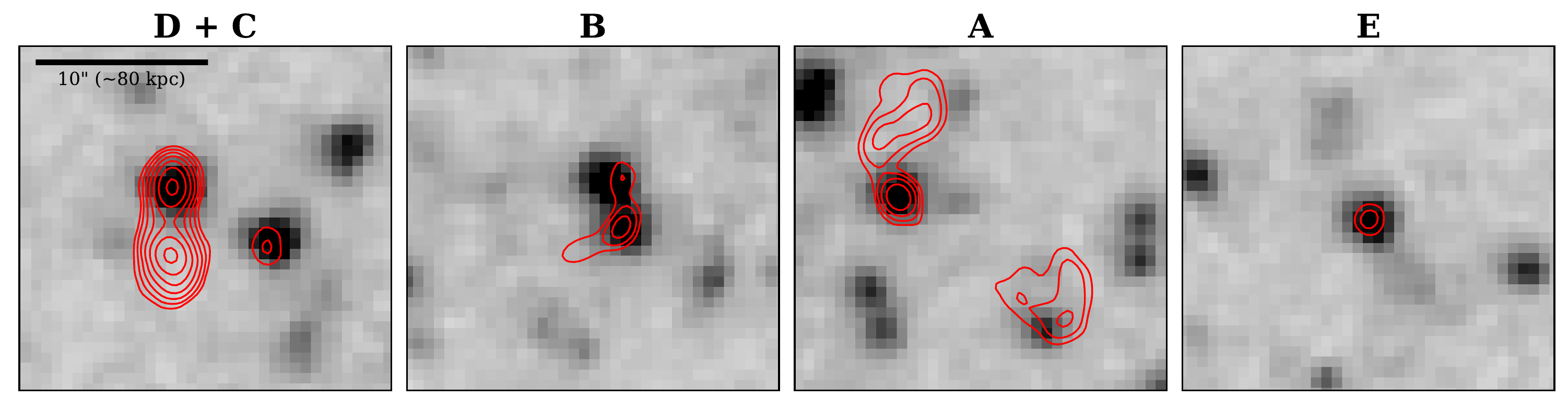}
	\caption{For each radio source, 20\as\ $\times$ 20\as\ 3.6 $\mu$m image overlaid with 1.4 GHz contours of the radio sources which start at 4$\sigma$ and increase by factors of 2$^n$ where n = 1,2,3, etc. The titles are the letter IDs referring to particular radio sources (see Table \ref{tb:rad_sources}). We identify counterparts for all radio sources except source D.}
	\label{fig:spz_cps}
\end{figure*}

\begin{deluxetable*}{cccccccc}
	\tablecaption{Radio Source Counterpart Properties \label{tb:cps}}
	\tablehead{\colhead{ID} & \colhead{Symbol} & \colhead{R.A.} & \colhead{Dec.} & \colhead{[3.6]}& \colhead{[4.5]} & \colhead{$r$} & \colhead{$z$}\\
	\colhead{} & \colhead{} & \colhead{(J2000)} & \colhead{(J2000)}& \colhead{(Vega)} & \colhead{(Vega)} & \colhead{(AB)} & \colhead{(AB)}}
	\startdata
     A &  $\mathlarger{\mathlarger{\bigstar}}$ &  15:06:20.41 &  +51:36:54.2 &  16.32$\pm$0.03 &  16.04$\pm$0.03 &   23.4$\pm$0.19 &  21.47$\pm$0.06 \\
     B &                      $\pentagonblack$ &  15:06:21.81 &  +51:36:44.5 &  17.35$\pm$0.07 &  17.14$\pm$0.09 &  24.25$\pm$0.47 &  21.61$\pm$0.07 \\
     C &                      $\mdlgblksquare$ &  15:06:24.78 &  +51:36:49.2 &  16.18$\pm$0.03 &  15.98$\pm$0.03 &  23.53$\pm$0.22 &  21.11$\pm$0.04 \\
     D & \ldots$^*$ & \ldots & \ldots & \ldots & \ldots & \ldots & \ldots \\
     E &                             \ding{58} &  15:06:15.92 &  +51:36:38.5 &  16.36$\pm$0.03 &  16.05$\pm$0.03 &  23.14$\pm$0.15 &  21.57$\pm$0.06 \\
	\enddata
	\tablecomments{ID is the letter ID of the corresponding radio source (see Figure \ref{fig:radio_data} and Table \ref{tb:rad_sources}). The symbols are those used to represent the counterpart in the figures of this work. The R.A. and Dec. are the coordinates of the \spz\ counterpart. [3.6] and [4.5] are the \spz\ channel 1 and channel 2 magnitudes of the counterpart in Vega magnitudes. The $r$ and $z$-band counterpart magnitudes are from \dc\ and are in AB magnitudes. $^*$No counterpart for radio source D is detected in either \spz\ or \dc.\\}
\end{deluxetable*}
%     D & $-^*$ & $-$ & $-$ & $-$ & $-$ & $-$ & $-$ \\

\begin{figure}
    \centering
	\includegraphics[width=\linewidth, keepaspectratio]{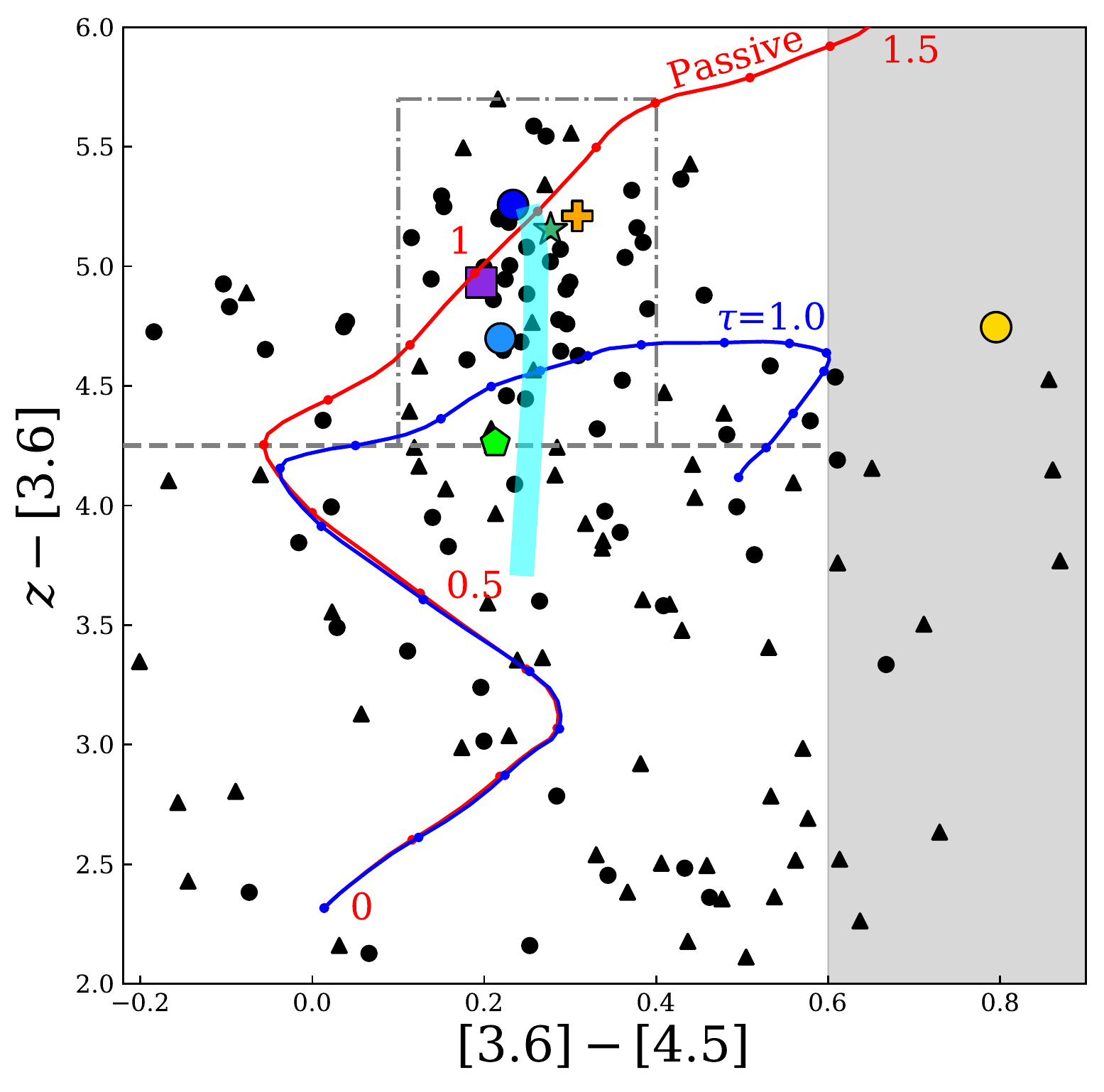}
	\caption{Color-color diagram of all sources within 1\am\ of the cluster center. The triangles indicate counterparts that only have optical lower limits. The radio source counterparts are marked according to Table \ref{tb:cps} where the dark green star is associated with the counterpart of radio source A, the light green pentagon with B, the purple square with C, and the yellow plus with E. The three brightest galaxies in the cluster are marked as light blue, dark blue, and yellow circles in order of decreasing brightness (see Table \ref{tb:bcgs} and \S\ref{sect:bcgs}). The red curve indicates the expected color evolution of a passive galaxy as a function of redshift, with each point marking the color at intervals of redshift $\delta z$ = 0.1. The blue curve indicates the expected color evolution of a galaxy with a star formation burst at the formation redshift of $z_f$ = 3.0 exponentially decaying over 1.0 Gyr. The thin vertical cyan shaded track marks the evolution of a galaxy at the cluster redshift ($z$=1.09) from a passive galaxy to a galaxy with a burst of star formation decaying over 0.1 Gyr, 0.5 Gyr, 1.0 Gyr, and 10.0 Gyrs. The horizontal gray dashed line is the expected $z-[3.6]$ color of a passive galaxy at $z = 0.7$. The shaded gray region represents color space in which the emission is generally dominated by emission from the AGN. Galaxies above the gray dashed line and to the left of the gray shaded region are considered candidate cluster members, while the subset of those points in the dashed-dotted box were used for the density analysis described in \S\ref{sect:merger}. All of the radio source counterparts are consistent with being a passive galaxy at $z>0.7$.}
	\label{fig:color_color_prog}
\end{figure}

\begin{figure}
    \centering
	\includegraphics[width=\linewidth, keepaspectratio]{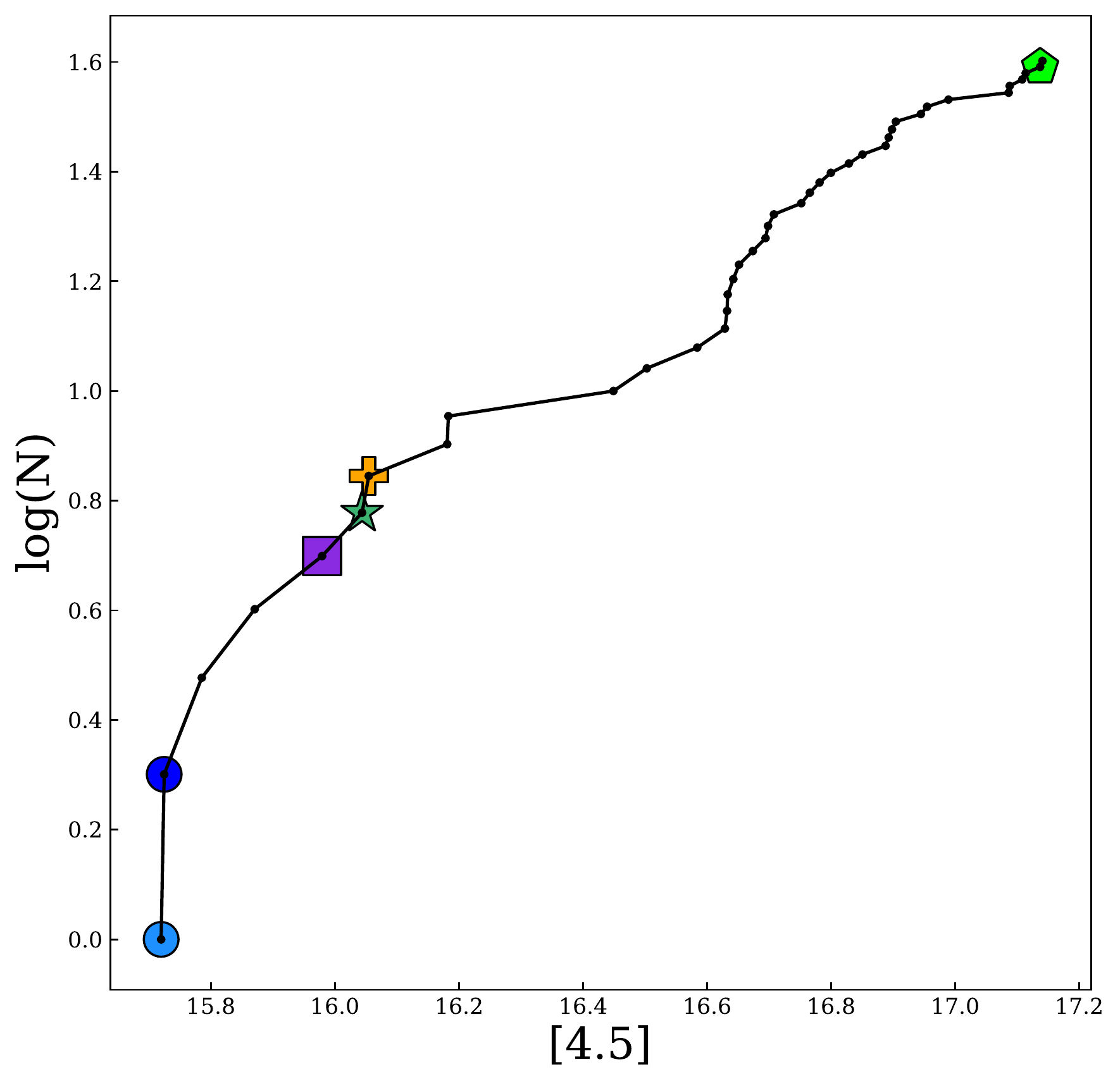}
	\caption{The cumulative histogram for the 40 brightest sources in [4.5] within 1\am\ of the cluster center that have colors consistent with being at $z>0.7$. The [4.5] values at which the radio source counterparts are marked by their respective symbols (see Table \ref{tb:cps} and Figure \ref{fig:color_color_prog}). Three of the seven most massive galaxies have associated radio emission.\\}
	\label{fig:prog_hist}
\end{figure}

Source A could also be classified as a HyMoRS due to its FR I appearance in the northern jet and its FR II appearance of the southern jet. However, we acknowledge that a HyMoRS morphology may arise due to projection effects \citep{Harwood20}. Additionally, there is an asymmetry in the length of the northern jet versus the southern jet with each being bent by roughly the same amount, but the southern jet extends further along the inferred jet axis than the northern jet.

The source to the east of the central radio source (marked as source B) is classified as a bent-tail. Showcased by the 1.4 GHz data, the jets are symmetrically bent. This symmetrical behavior is expected for a classical radio source in a uniform medium \citep{Falle91,Kaiser97}. The angle that the bent tails subtend is quite large and thus it is not classified as a narrow-angle tail or head-tail radio galaxy. Distortion of radio galaxy jets is a product of interaction with the ICM through ram pressure \citep{Begelman79}. The rather large opening angle demonstrated by this source thus indicates a lower velocity. The motion of this galaxy could be due to the merger or simply infalling due to typical motions of galaxies near the cluster center. 

The most western source (marked as source D) is an FR II source. It has the characteristic evenly spaced lobes and does not exhibit any bent characteristics. \cite{Garon19} find a similar trend that sources are bent more drastically closer to the cluster center and eventually become straight farther out in the cluster.

There are two radio point sources within 1\am\ of the cluster center: one to the east side of the cluster (marked as source C) near the FR II and one to the south east of the cluster (marked as source E).

Through our observations we also detected strong emission from a low redshift source to the west of the cluster (see Figure \ref{fig:source_F} in the Appendix). Since this source is not at the redshift of the cluster, it is not included in the main analysis of this work and we refer the reader to the Appendix for further details.

\section{Radio Source Counterparts Identification}\label{sect:cps}
We use \spz\ 3.6 and 4.5 $\mu$m data and \dc\ $r$ and $z$ data to identify candidate cluster members and analyze the counterparts of the radio sources (see Figure \ref{fig:spz_cps}). We cross match the \spz\ source catalog to the \dc\ catalog with a 1\as\ matching radius. If there is no match, we use the \dc\ 5$\sigma$ $r$ depth of 23.4 and $z$ depth of 22.5 to define the limit on the colors. In Figure \ref{fig:color_color_prog}, we plot all the \spz-\dc\ matched sources within 1\am\ of the cluster center (black points). 

To identify each radio source counterpart, we use the catalog of infrared sources to determine the closest \spz\ object within 1\as\ of the inferred origin of the radio emission. We find matches for four of the five radio sources and record their magnitudes and colors in Table \ref{tb:cps}. We note that the nearest match for the FR II radio galaxy (source D in Figure \ref{fig:spz_cps}) is 2\farcs1 from the inferred origin of the radio emission and is associated with the lobe instead of the origin of emission. We thus conclude that we do not detect a counterpart for source D. 

With the infrared and optical counterparts identified, we compare their observed colors to the expected colors of a passively evolving galaxy as a function of redshift using \verb'EzGal' \citep[]{EzGal12}. Following the parameters used in \cite{Gonzalez19} for \madcows, we use an FSPS model \citep[]{c09} with a simple stellar population, a \cite{Chab03} initial mass function (IMF), and a formation redshift of $z_f$ = 3. Because the brightest galaxies in clusters have super-solar metallicities \citep{Trager08,Connor19}, we also assume a chemical composition of $Z=0.03$. This model is plotted as a function of redshift in Figure \ref{fig:color_color_prog} (red curve). We also compare the counterpart colors to the expected colors of a galaxy with a 1.0 Gyr exponentially decaying star formation burst at the formation redshift ($z_f$=3.0) shown in Figure \ref{fig:color_color_prog} (blue curve). We use $z-$[3.6] instead of $r-$[3.6] to isolate cluster galaxy members because $z$ is deeper than $r$ and it has been shown that $z-$[3.6] increases proportionally to redshift \citep{Muzzin13}. We consider cluster members to be those that have colors consistent with being a passive galaxy at $z>0.7$ (colors above the gray dashed line in Figure \ref{fig:color_color_prog}; see \citealt{Gonzalez19} for an explanation of the method). 
In Figure \ref{fig:color_color_prog}, we see a clear overdensity corresponding to the passive evolutionary curve (red curve) at $z\sim1$. We identify the radio source counterparts with colored shapes (as indicated in Table \ref{tb:cps}) and compare them to the passive evolutionary track to identify any low redshift interlopers (below the dashed gray line which is the color expected for a passive galaxy at $z$ = 0.7) and those with AGN-dominated emission (shown by the shaded region; \citealt{Stern12}).

When comparing the colors of the host galaxies to the evolutionary track, we find that three of the counterparts are consistent with being a passive galaxy at $z>0.7$. The last counterpart (source B, lime green pentagon) is consistent with being a star forming galaxy at $z>0.7$. We find that none of the counterparts' emission is dominated by an AGN. 

In Figure \ref{fig:prog_hist}, we plot a cumulative histogram of the 40 brightest sources in [4.5] that have colors consistent with having a redshift $z\gtrsim0.7$ within 1\am\ of the cluster center. Under the assumptions that (1) these counterparts are cluster members and (2) the emission of the AGN is sub-dominant to that of the galaxy at these mid-infrared wavelengths, we can use IRAC [4.5] photometry as a proxy for stellar masses \citep{Eisenhardt08}. From this plot, we find that out of the highest stellar mass galaxies, three of these have radio emission. This is consistent with the findings of \cite{Moravec20} that radio source counterparts are among the most massive within the cluster environment.

\begin{deluxetable*}{cccccccc}
	\tablecaption{Brightest Galaxies in Cluster \label{tb:bcgs}}
	\tablehead{\colhead{Symbol Color} & \colhead{R.A.} & \colhead{Dec.} & \colhead{R$_{\mathrm{cc}}$}& \colhead{[3.6]}& \colhead{[4.5]} & \colhead{$r$} & \colhead{$z$}\\
	\colhead{} & \colhead{(J2000)} & \colhead{(J2000)} & \colhead{(\as)} & \colhead{(Vega)} & \colhead{(Vega)} & \colhead{(AB)} & \colhead{(AB)}}
	\startdata
     light blue &  15:06:24.75 &  +51:37:31.7 &  49.0 &  15.94$\pm$0.03 &  15.719$\pm$0.03 &  22.84$\pm$0.13 &  20.64$\pm$0.03 \\
       blue &  15:06:25.40 &  +51:36:52.1 &  45.0 &  15.96$\pm$0.03 &  15.724$\pm$0.03 &  23.44$\pm$0.22 &  21.21$\pm$0.05 \\
       yellow &  15:06:21.90 &  +51:36:47.0 &  18.0 &  16.03$\pm$0.03 &  15.232$\pm$0.03 &  22.66$\pm$0.18 &  20.77$\pm$0.06 \\
	\enddata
	\tablecomments{The symbol color listed is the color of the corresponding circle used in Figures \ref{fig:color_color_prog} and \ref{fig:spz_bcgs} to represent each of the brightest galaxies in the cluster. The R.A. and Dec. are the coordinates of the galaxy in \spz. R$_{\mathrm{cc}}$ is the distance from the \textit{WISE} center. [3.5] and [4.5] are the \spz\ channel 1 and channel 2 magnitudes (Vega). The $r$ and $z$-band magnitudes (AB) are from \dc. $^*$This is the brightest galaxy in [4.5] but its [3.6] $-$ [4.5] color indicates that its emission is dominated by an AGN (see \S\ref{sect:bcgs}).}
\end{deluxetable*}
%     D & $-^*$ & $-$ & $-$ & $-$ & $-$ & $-$ & $-$ \\

\section{Brightest Galaxies in \moo15}\label{sect:bcgs}
Brightest cluster galaxies are the most massive galaxies in their host cluster, and as a population are the most massive galaxies in the Universe at a given epoch. BCGs are typically elliptical galaxies which lie near the bottom of the cluster potential well.

We identify the BCG candidate with the following approach. We isolate the brightest source in [4.5] that is not dominated by AGN emission. If [3.6] $-$ [4.5] $>$ 0.6 this indicates that the emission is dominated by an AGN \citep{Stern12} and we are not able to obtain a stellar mass using [4.5] or robust photometric redshift. Thus, using the catalog of \spz-\dc\ cross-matched sources within 1\am\ of the cluster center, we identify galaxies that are not dominated by AGN emission by applying a color cut of [3.6] $-$ [4.5] $<$ 0.6. We then choose cluster members by requiring the $z-$[3.6] color to be consistent with the color expected for a passive galaxy at $z$ = 0.7 ($z-$[3.6] $>$ 4.25; calculated using \verb'EzGal'). Lastly, we use [4.5] as a proxy for stellar mass \citep{Eisenhardt08} and identify the most massive galaxies (the brightest sources in [4.5]) within 1\am\ of the cluster center.

\begin{figure}
    \centering
	\includegraphics[width=\linewidth, keepaspectratio]{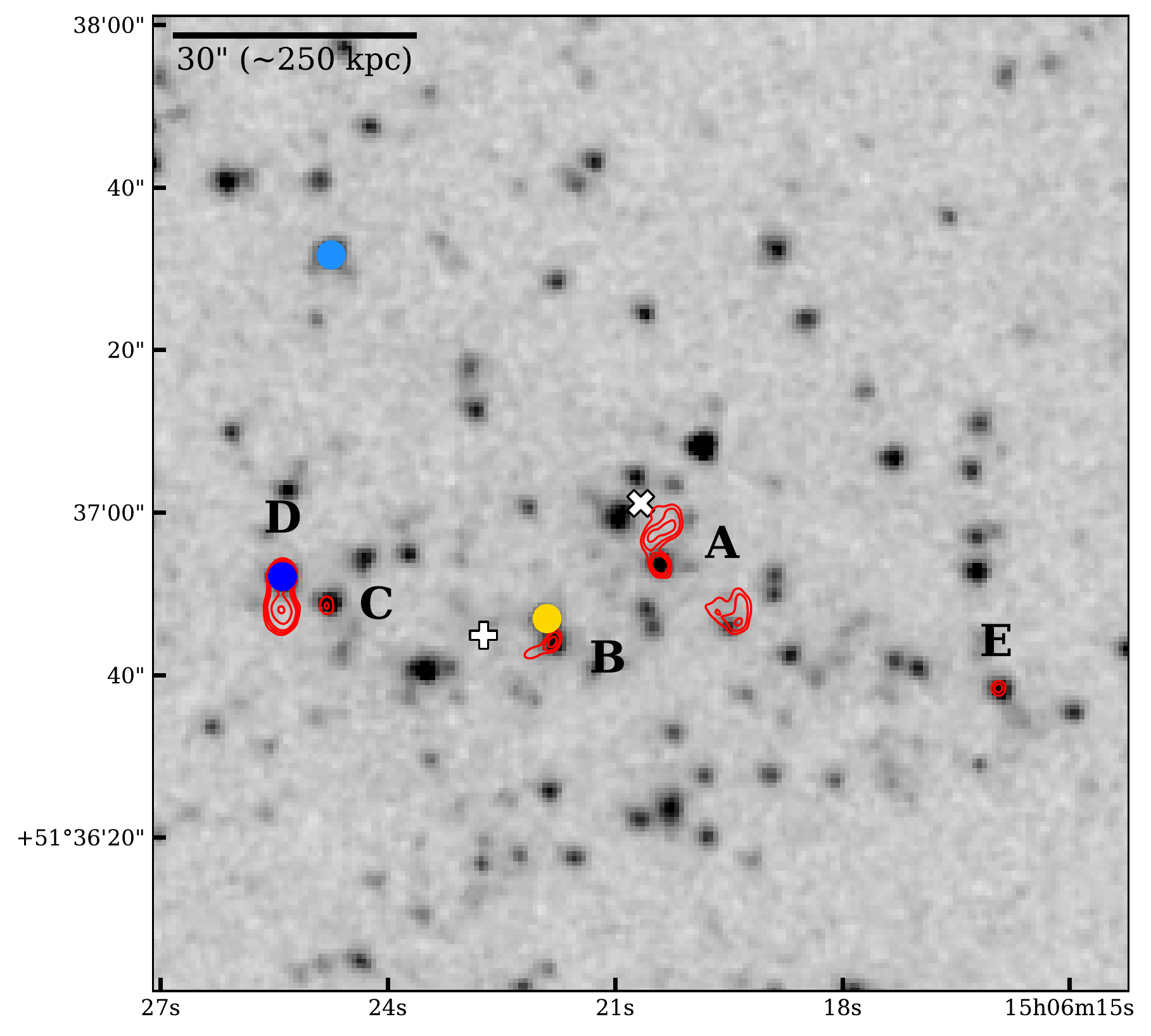}
	\caption{3.0\am\ $\times$ 3.0\am\ 3.6 $\mu$m image overlaid with VLA 1.4 GHz contours which start at 4$\sigma$ and increase by factors of 2$^n$ where n = 1,2,3, etc. The letter IDs are used to refer to particular radio sources (see Table \ref{tb:rad_sources}). The colored circles represent some of the brightest galaxies in the cluster (see Table \ref{tb:bcgs}). The SZ center is denoted by a white + and the \madcows\ \textit{WISE} center is denoted by a white x. North is up and east is to the left.}
	\label{fig:spz_bcgs}
\end{figure}

We find two galaxies that are equal in stellar mass within the uncertainties ($\delta_{[4.5]}$ = 0.005 see Table \ref{tb:bcgs}). One of these galaxies is 45\as\ of the cluster center (dark blue), while the other is offset 49\as\ northeast of the cluster center (light blue). We consider these two galaxies the most probable candidate BCGs in the system. There is another potential BCG candidate 18\as\ east of the cluster center that is brighter in [4.5] than both other candidates, but it is an AGN-dominated galaxy and thus we cannot estimate its stellar mass or redshift  (yellow circle in Figures \ref{fig:color_color_prog} and \ref{fig:spz_bcgs}). We note that we do not associate the BCG candidate near the cluster center (dark blue) with the radio emission of the nearby FR II, as there is a spatial offset of 2\farcs1 from the inferred origin of the radio emission. Thus, we have no compelling evidence for radio emission associated with any of the BCG candidates.

\section{Evidence for Merger}\label{sect:merger}
Previous studies have found that a merging cluster can have enhanced AGN and radio-AGN activity \citep{Owen99,Miller03,Sobral15}. There are numerous methods to determine whether this cluster is merging with another structure. Below we describe three methods that produce three pieces of evidence that \moo15\ is a merging system. 

\subsection{Spatial Density of High Redshift Galaxies}\label{sect:density}
First we investigate the density and distribution of sources at the cluster redshift. We use a similar method to that described in \S\ref{sect:cps} to isolate galaxies at the cluster redshift by applying a series of color cuts. Using the catalog of \spz-\dc\ cross-matched sources across the entire field over which we have \spz\ data (within 4\am\ of the cluster center), we discard any whose emission is dominated by an AGN (i.e., [3.6]$-$[4.5] $>$ 0.6). 

\begin{figure*}
    \centering
	\includegraphics[width=\linewidth, keepaspectratio]{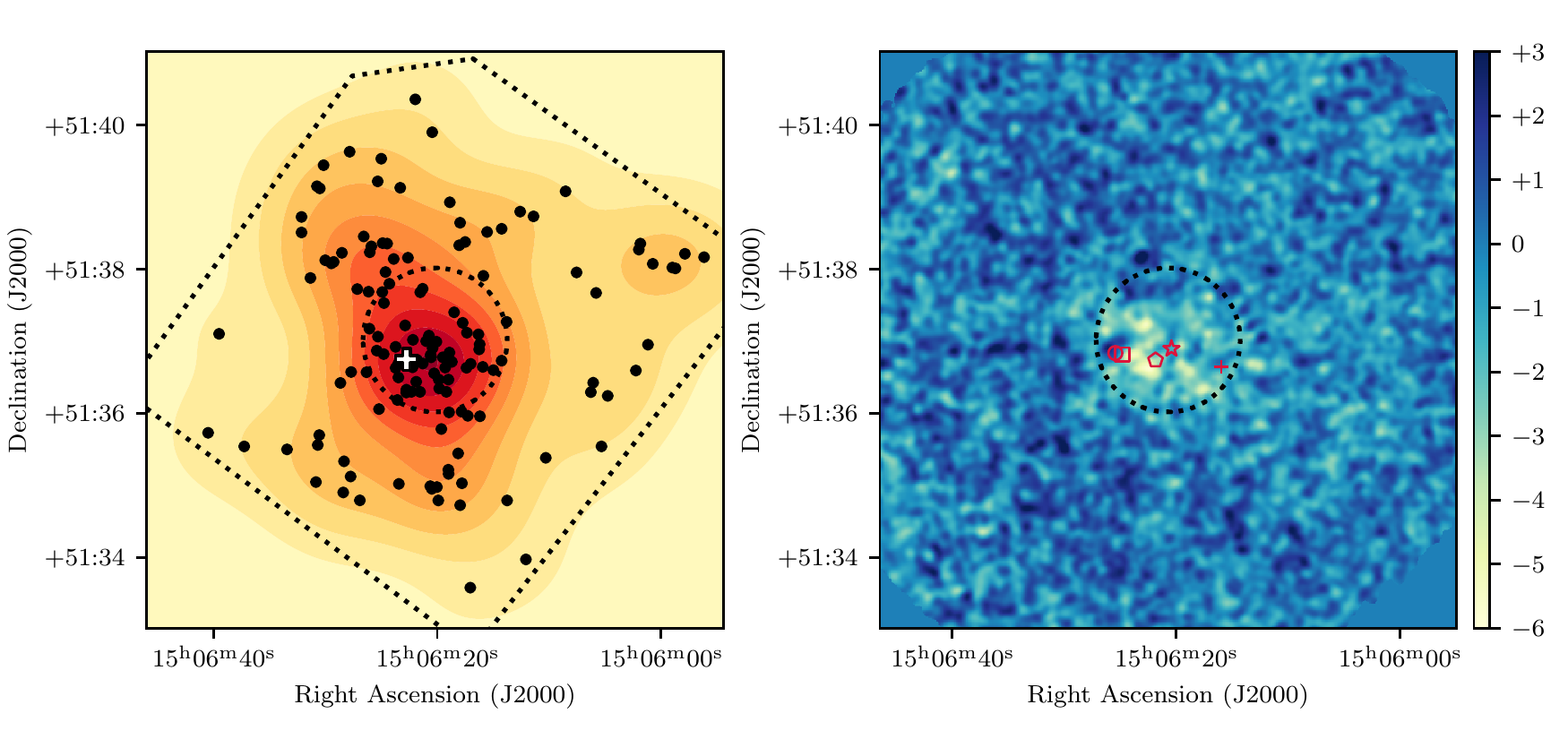}
	\caption{\textbf{Left:} density plot of the points (black) that are the \spz\ sources that are consistent with being a passive galaxy in the cluster or a star forming galaxy with a star formation burst of 1.0 Gyr (see the gray dashed box in Figure \ref{fig:color_color_prog} and \S\ref{sect:density}). The contours show density levels as a result of a KDE smoothing kernel using Silverman's method to determine kernel size. The black dashed circle is a 1\am\ circle centered on the cluster center from \madcows. The \spz\ image boundary and thus catalog limits are shown by black dashed lines. The SZ center is marked with a white +. There is evidence for a substructure to the northeast of the cluster (see \S\ref{sect:density}). \textbf{Right:} 90 GHz signal-to-noise map with levels corresponding to the color bar to the right. The radio source locations are marked with open shapes according to their counterpart (see Table \ref{tb:cps}). There is no counterpart for source D, thus we mark the radio source location with an open circle.}
	\label{fig:sub_structure}
\end{figure*}

We identify the galaxies that have the highest probability of being cluster members with a method similar to that of \cite{Gonzalez19}. As \cite{Gonzalez19}, we require that a galaxy have a [3.6]$-$[4.5] color within $\pm$ 0.15 magnitudes of that expected from a model of a passively evolving galaxy at the cluster redshift, noting that passive and star-forming galaxies have similar [3.6]$-$[4.5] colors at this epoch (see gray dashed box in Figure \ref{fig:color_color_prog}). The width of the color window in [3.6]$-$[4.5] is designed to be inclusive of likely cluster members given the photometric uncertainties, while still enabling a meaningful reduction of the background contribution. For the $z-$[3.6] color, we choose limits that retain sources that are relatively consistent with both the passive galaxy (red curve in Figure \ref{fig:color_color_prog}) and a star forming galaxy with a burst lasting 1.0 Gyr (blue curve in Figure \ref{fig:color_color_prog}) at $z>0.7$. These color criteria are illustrated by the dashed-dotted box in Figure \ref{fig:color_color_prog}.

The spatial distribution of the sources that meet these color criteria is shown in Figure \ref{fig:sub_structure}. The contours show density levels as a result of a kernel density estimation (KDE) smoothing kernel using Silverman's method to determine kernel size (using \verb'scipy'). The main cluster can be clearly seen in the contours at the center of Figure \ref{fig:sub_structure}. To the northeast of the cluster, there is evidence of a substructure, and we see evidence of elongation in the galaxy distribution along the axis connecting this substructure and the main cluster.

\begin{figure}
    \centering
	\includegraphics[width=\linewidth, keepaspectratio]{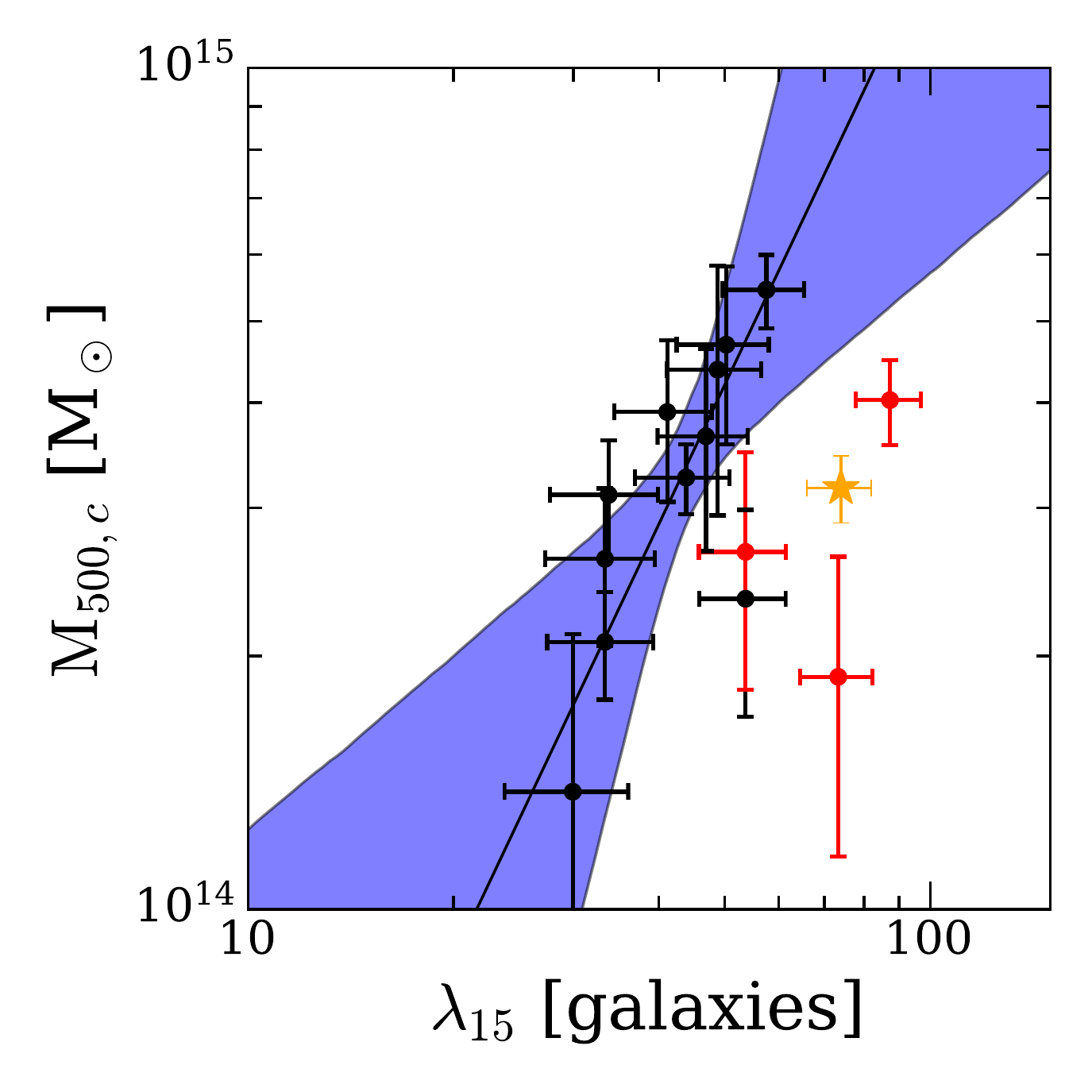}
	\caption{SZ-based M$_{500}$ versus $\lambda_{15}$ (richness) for \madcows\ clusters. Systems that are known mergers based upon \textit{Chandra} observations are denoted as red points and a best-fit relation is derived excluding these clusters. The best-fit relation is shown as a solid black line, while the shaded region denotes the 68\% confidence interval (see \citealt{Gonzalez19} for details). Using the SZ mass from the GBT data, MOO J1506+5137 is marked with a yellow star and its position indicates that it is a merging system.}
	\label{fig:mass_richness}
\end{figure}

\subsection{Radio Morphology of Jets}
As mentioned before, there are five radio sources in MOO J1506+5137; two of which are distorted and bent. Distortion of radio galaxy jets is a product of interaction with the ICM \citep{Begelman79}. NATs are a result of ram pressure sweeping the jets back as the host galaxy moves at high speed through the ICM \citep{Miley72,Begelman79}. However, it has been proposed and corroborated that instead the bending of a WAT is an indication of large-scale bulk motions in the ICM and is thus indicative of a cluster merger \citep{Roettiger93,Pinkney94,Roettiger96,Burns98,Sakelliou00,Burns02}. We detect 2 BTs, one of which is classified as a possible WAT/HyMoRS whose morphology is therefore suggestive of a cluster merger. 

Concerning the HyMoRS classification of the central source, a recent study of 25 HyMoRS found two that were bent and one that was in a galaxy cluster \citep{Kapinska17}. One formation theory of HyMoRS is that the different jets are produced based upon jet interaction with the external medium \citep{Gopal-Krishna00}. As suggested by \cite{Gawronski06}, if there is reason to believe that the conditions and properties of the ICM in a galaxy cluster differ between the two opposite sides of the radio galaxy, and the jets are identically launched, then a HyMoRS could emerge. The HyMoRS characteristics of the BT source further corroborate the emerging picture that \moo15\ is a merging system. However, as noted previously, we acknowledge that the HyMoRS morphology classification arise due to projection effects \citep{Harwood20}.

Further, there is an asymmetry in the length of the northern jet versus the southern jet with each being bent by roughly the same amount, but the southern jet extends further along the inferred jet axis than the northern jet. One could imagine this asymmetry arising if there is a significant ICM gradient due to a merger which would shorten the length of the jet in the direction of the merger. However, this morphology may be an effect of projection with respect to our point of view.

\subsection{Mass-Richness Relation}\label{sect:mass-rich}
In the \madcows\ survey, $\lambda_{15}$ is defined as the richness within a 1 Mpc aperture centered on the cluster above a flux density threshold of \spz\ 4.5 $\mu$m 15$\mu$Jy \citep{Gonzalez19}. Whenever available, we compare the SZ-determined masses to the richness estimate for each \madcows\ cluster, and a mass-richness relation emerges (see Figure \ref{fig:mass_richness}). The clusters that lie off to the right of the relation are known mergers based upon \textit{Chandra} observations and are denoted as red points. We expect merging systems to lie to the right of the mass-richness relation because when two systems come together we will observe an enhancement in the number of galaxies in the system before the gas from the two subclusters coalesces and the SZ signal increases \citep{Poole07,Krause12,Yu15}.

We use SZ-determined mass from the GBT data and the richness from the \spz\ data for \moo15\ to place the cluster on the known \madcows\ mass-richness relation (marked as a yellow star in Figure \ref{fig:mass_richness}). MOO J1506+5137 lies in the regime with the merging systems, suggesting that MOO J1506+5137 is also a merging system. 

\subsection{Summary}
In summary, we find three pieces of evidence that lead to the conclusion that this is a merging system. First, through analysis of the density and spatial distribution of galaxies around the cluster redshift, we find evidence of a subgroup to the northeast of the main cluster. Second, there are two centrally located bent-tail radio sources, one of which is a possible WAT/HyMoRS which can indicate merger activity. Third, the SZ mass and richness of this cluster place it in the regime of merging clusters on the \madcows\ mass-richness relation.

\section{Discussion}\label{sect:discuss}
Previous studies have also found an excess of radio galaxies in merging clusters \citep{Owen99,Miller03}. Cluster mergers are thought to affect radio emission in clusters on a wide range of scales: from the smaller scale of bent-tail galaxies to the larger scale of radio halos and relics. In this work, we find two bent-tail sources which indicate complicated dynamics within the cluster. Similar to \cite{Miller03}, we are led to the notion that the dynamical state of this merging system is somehow responsible for the relatively large  number and interesting morphologies of the radio sources. Physical processes that could plausibly be responsible for heightened radio activity in a merging system include perturbation of the cool gas disk within the cluster galaxies via interactions or ram pressure effects, and gas-dynamical processes. 

In \S\ref{sect:cps}, we established that the host galaxies are consistent with being passive (presumably elliptical) galaxies. Nearby radio galaxies are known to often contain gas disks \citep{Martel00}. Strong galaxy-galaxy and galaxy-cluster potential interactions that are common within galaxy clusters could perturb a cool gas disk in cluster galaxies. Due to the high speeds of galaxies in galaxy clusters, cluster galaxies are likely to experience repeated close encounters with neighbors or galaxy-galaxy interactions \citep{Moore96,Moore99} which could perturb the gas disk. Also in the core of a galaxy cluster, tidal interactions between cluster galaxies and the cluster potential can affect the structure of a galaxy and the distribution of the gas inside it \citep{Fujita98,Natarajan98}. Thus, perturbation of the gas disk within the elliptical hosts is a viable option to explain the heightened radio activity in this merging cluster. This warrants investigations with simulations that can trace the impact on these disks as a central factor to the radio feedback cycle in clusters.

Gas-dynamical processes could also cause heightened radio activity in a merging system. For example, for a galaxy passing through a merger shock, the ambient pressure would increase, possibly leading to the compression and cooling of gas in the galaxy, which would create fuel for the AGN \citep{Sobral15}. Ram pressure and thermal conduction could also affect the central gas in a galaxy. For instance, \citet{Poggianti17} find evidence that galaxies in the midst of ram pressure stripping host AGN and \citet{Nipoti07} find that thermal conduction affects the properties of the gas that is accreted onto the black hole.

We note that these results are consistent with the heightened AGN and radio activity in the center of clusters at high redshift \citep[][Mo et al., sub.]{Galametz09,Martini13,Wylezalek13,Alberts16,Bufanda17,Mo18}. This work could indicate that the heightened activity is due to the merging of clusters which is common at the epoch of $1<z<2$ as clusters are rapidly assembling and forming through mergers \citep{Bode01}. From this work, \cite{Owen99}, and \cite{Miller99}, we see that mergers are associated with radio-AGN activity which will be important for understanding the development and energetics of the ICM as the cluster evolves.

\section{Summary}\label{summary}
We present the multi-wavelength analysis of a highly radio-active galaxy cluster from the Massive and Distant Clusters of \wise\ Survey (\madcows), MOO J1506+5137. The multiwavelength dataset is rich, comprised of 144 MHz observations from LoTSS, 1.4 GHz and 6.0 GHz VLA observations, 90 GHz MUSTANG-2 observations, \spz\ 3.6 and 4.5 $\mu$m observations, and $r$ and $z$ observations from \dc. Below is a summary of our findings:
\begin{itemize}
    \item \textit{Radio Sources:} MOO J1506+5137 is an unusually radio-active cluster compared to other \madcows\ clusters. It contains five radio sources, of which three are complex. The radio morphologies include one BT/WAT/HyMoRS, one BT, one FR II, and two point sources.
    \item \textit{Counterparts/Host Galaxies:} We identified counterparts for four out of the five radio sources. Three of these counterparts have colors consistent with a passive galaxy at $z>0.7$ and one has colors consistent with a star forming galaxy at $z>0.7$. Out of the seven galaxies with the highest stellar mass in the cluster, three of them have associated radio emission and none of these three are the BCG candidates.
    \item \textit{Evidence for Merging System:} There is compelling evidence that this cluster is a merging system. First, the spatial density and distribution of galaxies approximately at the cluster redshift indicate that there is a subgroup to the northeast of the main cluster. Second, the presence of two BT sources -- one of which is a possible WAT/HyMoRS -- indicate a merging and chaotic environment. Third, the SZ mass and richness of this cluster place it in the regime of merging clusters on the \madcows\ mass-richness relation. 
\end{itemize}

These data are consistent with previous studies that have also found an excess of radio activity in galaxies in merging clusters \citep[e.g.,][]{Owen99,Miller03}. These data are suggestive that \textit{during the merger phase radio activity can be dramatically enhanced}, which would contribute to the observed trend of increased radio activity in clusters with increasing redshift.

\begin{figure*}
    \centering
	\includegraphics[width=\linewidth, keepaspectratio]{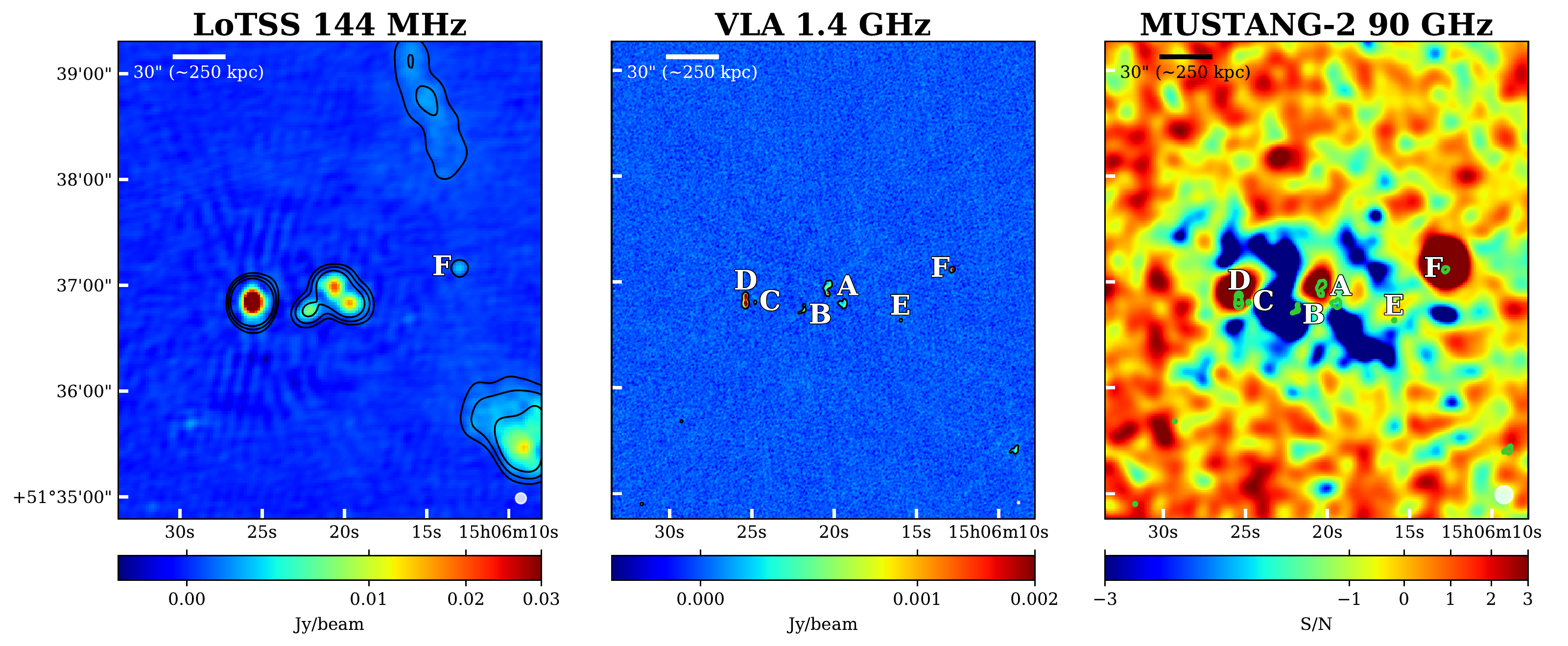}
	\caption{4.0\am\ $\times$ 4.0\am\ image centered on MOO J1506+5137 that showcases a low redshift source to the west of the cluster marked as source F. \textbf{Left:} the 144 MHz LoTSS data with contour levels 4$\sigma$, 8$\sigma$, and 16$\sigma$. The contours are smoothed by a Gaussian 3x3 pixel smoothing kernel. \textbf{Middle:} the 1.4 GHz (L-band) VLA data with a contour level of 4$\sigma$ shown in black. The source labels correspond to the radio sources in Table \ref{tb:rad_sources}. \textbf{Right:} Green 1.4 GHz VLA contours (4$\sigma$ and 256$\sigma$) overlaid on 90 GHz \m2\ data. The synthesized beam size is shown in the lower right hand corner. North is up and east is to the left.}
	\label{fig:source_F}
\end{figure*}

\acknowledgements
E.M. would like to thank those at the NRAO workshops and helpdesk for invaluable help with imaging. E.M. would like to thank the anonymous referee for their helpful and insightful comments.

This research was supported in part by the National Science Foundation (AST-1715181). E.M. acknowledges financial support from the Czech Science Foundation project No.19-05599Y. This work was supported by the EU-ARC.CZ Large Research Infrastructure grant project LM2018106 of the Ministry of Education, Youth and Sports of the Czech Republic. This research was also supported in part by NASA through the NASA Astrophysical Data Analysis Program, award NNX12AE15G. Parts of this work have also been supported through NASA grants associated with the \textit{Spitzer} observations (PID 90177 and PID 11080). Basic research in radio astronomy at the U.S. Naval Research Laboratory is supported by 6.1 Basic Research. The work of D.S., P.E., and T.C. was carried out at the Jet Propulsion Laboratory, California Institute of Technology, under a contract with NASA.

This publication makes use of Common Astronomy Software Applications (CASA) package \citep[]{casa} and Astropy, a community-developed core Python package for Astronomy \citep[]{astropy} and APLpy, an open-host plotting package for Python \citep[]{aplpy}, and SciPy \citep{scipy}.

This publication makes use of data products from NSF's Karl G. Jansky Very Large Array (VLA). The National Radio Astronomy Observatory is a facility of the National Science Foundation operated under cooperative agreement by Associated Universities, Inc. This work was supported by the Green Bank Observatory which is a major facility funded by the National Science Foundation operated by Associated Universities, Inc. MUSTANG-2 is supported by the NSF award number 1615604 and by the Mt.\ Cuba Astronomical Foundation.  Additionally, this publication makes use of data products from the \textit{Wide-field Infrared Survey Explorer}, a joint project of the University of California, Los Angeles, and the Jet Propulsion Laboratory/California Institute of Technology, funded by NASA. This work is also based in part on observations made with the \textit{Spitzer Space Telescope}, which is operated by the Jet Propulsion Laboratory, California Institute of Technology under a contract with NASA. 

\facilities{LOFAR (LoTSS), VLA, GBT (MUSTANG-2), \textit{Spitzer}, \textit{WISE}, Blanco (DECaLS)}
\software{APLpy, Astropy, Astroquery, CASA, EzGal, pandas, SciPy}
\newline

%%%%%%%%%%%%%%%%% APPENDICES %%%%%%%%%%%%%%%%%%%%%
\appendix
Through our observations we discovered a low redshift source to the west of the cluster (source F in Figure \ref{fig:source_F}). This source is a known low-redshift FR II source. At high frequencies the core is seen ($\alpha_{2000}$ = 15h06m12s.81, $\delta_{2000}$ = +51\deg37\arcmin7 3\as), but at low frequency in the LoTSS data, the full FR II structure can be seen. Using the technique described in \S\ref{sect:vla}, we calculate $S_{\mathrm{1.4GHz}}$ = 6.66$\pm$0.05 mJy and $S_{\mathrm{6.0GHz}}$ =  9.49$\pm$0.02 mJy. Using Eqn. \ref{eqn:power}, we obtain $P_{1.4}$ = 0.79$\pm$0.01 $\times$ 10$^{25}$ W Hz${-1}$ assuming $z$=0.611 (see below) and a spectral index typical of radio cores, $\alpha$ = -0.3 \citep{Morganti97,Yuan18}.

The core has an optical counterpart that is detected in SDSS (SDSS J150612.81+513707.0) and has a spectroscopic redshift of $z$=0.611\footnote{\url{http://skyserver.sdss.org/dr15/en/tools/explore/Summary.aspx?id=1237659324947431986}} \citep{SDSSDR15}. The SDSS spectrum shows an early-type spectrum with prominent Calcium, Hydrogen, and Potassium absorption and a 4000 \AA\ break. The spectrum also shows strong indications of a buried AGN with narrow [OII] and [OIII] emission lines, but no evident H$\beta$ emission. 

The host galaxy is detected in \spz\ (within 1\as\ of the radio core center) at $\alpha_{2000}$ = 15h06m12s.85, $\delta_{2000}$ = +51\deg37\arcmin7 6\as\ with [3.6] = 15.19$\pm$0.03 (Vega) and [4.5] = 15.08$\pm$0.03 (Vega). The \dc\ counterpart within 1\as\ of the \spz\ galaxy has  $r$ = 20.08$\pm$0.02 (AB) and $z$ = 18.61$\pm$0.01 (AB). Using the color-color analysis described previously, we find that the host galaxy is consistent with being a passive galaxy at low redshift which agrees with the early-type classification from the SDSS spectrum.
%%%%%%%%%%%%%%%%%%%%%%%%%%%%%%%%%%%%%%%%%%%%%%%%%%
%%%%%%%%%%%%%%%%%%%%%%%%%%%%%%%%%%%%%
\bibliography{bibliography_M1506}{}
\bibliographystyle{aasjournal}

\end{document}

%% file: authors.tex
%% 
%ORDER TBD
%%
\author[0000-0001-9793-5416]{Emily Moravec}
\affiliation{Astronomical Institute of the Czech Academy of Sciences, B\v ocn\'i II 1401/1A, 14000 Praha 4, Czech Republic}
\affiliation{Department of Astronomy, University of Florida, 211 Bryant Space Science Center, Gainesville, FL 32611, USA}
\author[0000-0002-0933-8601]{Anthony H. Gonzalez}
\affiliation{Department of Astronomy, University of Florida, 211 Bryant Space Science Center, Gainesville, FL 32611, USA}
\author[0000-0002-1940-4289]{Simon Dicker} 
\affiliation{Department of Physics and Astronomy, University of Pennsylvania, 209 South 33rd Street, Philadelphia, PA, 19104, USA}
%%%%%%%%%%%
\author[0000-0002-8909-8782]{Stacey Alberts}
\affiliation{Steward Observatory, 933 N Cherry Ave, Tucson, AZ 85721, USA}
\author[0000-0002-4208-798X]{Mark Brodwin}
\affiliation{Department of Physics and Astronomy, University of Missouri, 5110 Rockhill Road, Kansas City, MO 64110, USA}
\author[0000-0001-6812-7938]{Tracy E. Clarke}
\affiliation{Naval  Research  Laboratory, Code 7213,  Washington, DC 20375, USA}
\author[0000-0002-7898-7664]{Thomas Connor}
\affiliation{Jet Propulsion Laboratory, California Institute of Technology, 4800 Oak Grove Drive, Pasadena, CA 91109, USA}
\author{Bandon Decker}
\affiliation{Department of Physics and Astronomy, University of Missouri, 5110 Rockhill Road, Kansas City, MO 64110, USA}
\author[0000-0002-3169-9761]{Mark Devlin}
\affiliation{Department of Physics and Astronomy, University of Pennsylvania, 209 South 33rd Street, Philadelphia, PA, 19104, USA}
\author{Peter R. M. Eisenhardt}
\affiliation{Jet Propulsion Laboratory, California Institute of Technology, 4800 Oak Grove Drive, Pasadena, CA 91109, USA}
\author[0000-0002-8472-836X]{Brian S.\ Mason}
\affiliation{National Radio Astronomy Observatory, 520 Edgemont Rd., Charlottesville VA 22903, USA}
\author[0000-0002-9179-9801]{Wenli Mo}
\affiliation{Department of Astronomy, University of Florida, 211 Bryant Space Science Center, Gainesville, FL 32611, USA}
\author[0000-0003-3816-5372]{Tony Mroczkowski}
\affiliation{European Southern Observatory, Karl-Schwarzschild-Str.\ 2, D-85748 Garching b.\ M\"unchen, Germany}
\author[0000-0001-8592-2706]{Alexandra Pope}
\affiliation{Department of Astronomy, University of Massachusetts, 710 North Pleasant Street Amherst, MA 01003, USA}
\author[0000-0001-5725-0359]{Charles E.\ Romero}
\affiliation{Department of Physics and Astronomy, University of Pennsylvania, 209 South 33rd Street, Philadelphia, PA, 19104, USA}
\affiliation{Green Bank Observatory,  P.O. Box 2, Green Bank, WV 24944, USA}
\author[0000-0003-0167-0981]{Craig Sarazin} 
\affiliation{Department of Astronomy, University of Virginia, P.O. Box 400325, Charlottesville, VA 22901, USA}
\author[0000-0001-6903-5074]{Jonathan Sievers}
\affiliation{Department of Physics, McGill University, 3600 Rue University, Montr\'eal, QC H3A 2T8, Canada}
\affiliation{McGill Space Institute, McGill University, 3550 Rue University, Montr\'eal, QC H3A 2A7, Canada}
\affiliation{School of Chemistry and Physics, University of KwaZulu-Natal, Private Bag x54001, Durban 4001, South Africa}
\author{Spencer A. Stanford}
\affiliation{Department of Physics, University of California, Davis, One Shields Avenue, Davis, CA 95616, USA}
\author[0000-0003-2686-9241]{Daniel Stern}
\affiliation{Jet Propulsion Laboratory, California Institute of Technology, 4800 Oak Grove Drive, Pasadena, CA 91109, USA}
\author[0000-0003-2212-6045]{Dominika Wylezalek}
\affiliation{European Southern Observatory, Karl-Schwarzschild-Str.\ 2, D-85748 Garching b.\ M\"unchen, Germany}
\author[0000-0001-6527-828X]{Fernando Zago}
\affiliation{Department of Physics, McGill University, 3600 Rue University, Montr\'eal, QC H3A 2T8, Canada}
\affiliation{McGill Space Institute, McGill University, 3550 Rue University, Montr\'eal, QC H3A 2A7, Canada}
%% 

%%% MaDCoWS affiliated
% Alberts, Brodwin, Decker, Eisenhardt, Pope, Mo, Stanford, Stern, Wylezalek, Connor

%%% MUSTANG2 affiliated
% Dicker, Delvin, Romero, Mason, Mroczkowski, Sarazin, Sievers, Zago

%%% Other
% Clarke